\documentclass[prl,aps,preprint,bibtex,showpacs]{revtex4}
\usepackage{graphics}

\begin{document}

\title{On possible skewon effects on light propagation}

\author{Yuri N.~Obukhov\footnote{On leave from: Dept.\ of Theoret.\ 
Physics, Moscow State University, 117234 Moscow, Russia} and 
Friedrich W.~Hehl\footnote{Also at: Dept.\ of Phys.\ Astron., 
University of Missouri-Columbia, Columbia, MO 65211, USA}}
\affiliation{Institute for Theoretical Physics, University of Cologne,
50923 K\"oln, Germany}

%\date{19 November 2004, {\it file skewon5.tex}}
%\maketitle

\begin{abstract}
  We start from a {\it local\/} and {\it linear\/} spacetime relation
  between the electromagnetic excitation and the field strength. Then
  we study the generally covariant Fresnel surfaces for light rays and
  light waves. The metric and the connection of spacetime are left
  unspecified. Accordingly, our framework is ideally suited for a
  search of possible violations of the Lorentz symmetry in the photon
  sector of the extended standard model. We discuss how the skewon
  part of the constitutive tensor, if suitably parametrized,
  influences the Fresnel surfaces and disturbs the light cones of
  vacuum electrodynamics. Conditions are specified that yield the
  reduction of the original {\it quartic\/} Fresnel surface to the
  double light cone structure (birefringence) and to the single light
  cone. Qualitatively, the effects of the real skewon field can be
  compared to those in absorbing material media. In contrast,
  the imaginary skewon field can be interpreted in terms of
  non-absorbing media with natural optical activity and Faraday
  effects.  The astrophysical data on gamma-ray bursts are used for
  deriving an upper limit for the magnitude of the skewon field.
\end{abstract}
\pacs{PACS no.: 03.50.De, 41.20.Jb, 04.20.Gz}
\bigskip

\maketitle

%%%%%%%%%%%%%%%%%%%%%%%%%%%%%%%%%%%%%%%%%%%%%%%%%%%%%%%%%%%
\section{Introduction}
%%%%%%%%%%%%%%%%%%%%%%%%%%%%%%%%%%%%%%%%%%%%%%%%%%%%%%%%%%%

Classical electrodynamics is fundamentally linked to the concepts of
relativity and spacetime symmetry. In particular, many high precision
tests of relativity are based on electromagnetic phenomena.  At
present, there is a growing interest (both theoretically and
experimentally) in the search for possible {\it violations of the
  Lorentz symmetry}. The latter may naturally occur, for example, in
the framework of unified theories of physical interactions at the
Planck length scale, and may then manifest themselves at the scales of
usual high-energy particle physics in the form of corrections to the
established quantum field-theoretical models. Even though presently
the limits of the standard model of elementary particle physics (SM)
become visible, see Sec.10.6 of the new Particle Data Report
\cite{Cern}, it still remains the basis of our understanding of the
strong and the electroweak interactions. At the foundations of the SM
lays special relativity with the Poincar\'e symmetry group and,
associated with it, the {\it light cone\/} structure. Although this {\it
  rigid\/} light cone structure of special relativity is assumed as
framework for the SM, we know from general relativity theory (GR) and
from experiments
% (think of the influence of GR on the GPS, e.g.) 
that the light cone structure is in fact ``flexible''. In other words, 
in GR the metric becomes a space and time dependent field. This alone 
demonstrates that the SM cannot be strictly correct. 

Numerous attempts have been made to take care of this state of affairs
and to build up a more general framework for the SM and perhaps even
to include GR. One of the promising approaches is the standard model
extension (SME) of Kosteleck\'y and collaborators, see
\cite{ColladayKostelecky98,Kostelecky04,conf04}, in which by means of
a set of parameters certain violations of the SM are put in a
quantitative form amenable to experimental tests. The photon sector of
the SME has been particularly closely studied in
\cite{KosteleckyMewes02}, where the propagation of electromagnetic
waves in vacuum was investigated with Lorentz violating terms added in
the effective {\it constitutive\/} tensor.

In this paper, we continue the study of electrodynamical phenomena in
the framework of the so-called premetric approach, i.e., without
assuming a special form of the metric structure on a spacetime
manifold.  Such an approach provides a general technique for an
appropriate physical discussion of the experimental tests of possible
violations of the Lorentz symmetry and of parity, since at the outset
no assumptions were allowed about geometric structures, such as the
metric or the connection of spacetime. On the contrary, the analysis
of the dynamics of the electromagnetic field in this approach yields,
via the light propagation, the information about the underlying
metrical structure of spacetime.  Accordingly, this scheme can be
viewed as a general test theory of the searches for possible
deviations from the Lorentz symmetry in the photon sector of SME. It
is worthwhile to note that the multi-parameter approach of
Kosteleck\'y and Mewes \cite{KosteleckyMewes02} can be naturally
embedded into the framework of general premetric electrodynamics.

Electromagnetic wave propagation is a very important physical phenomenon 
in classical field theory. In general, the geometrical structure of 
spacetime as well as the intrinsic properties and the motion of material 
media can affect the light propagation. Hence, a theoretical analysis
of the latter ultimately results in establishing the properties of the
(genuine or effective ``optical") metric structure on the manifold.

In the generally covariant premetric approach to electrodynamics
\cite{Schouten,Post,HO02,Lindell}, the axioms of electric charge and
of magnetic flux conservation manifest themselves in the Maxwell
equations for the excitation $H=({\cal D},{\cal H})$ and the field
strength $F=(E,B)$:
\begin{equation}\label{me}
dH=J, \qquad dF=0 \,.
\end{equation}
These equations should be supplemented by a constitutive law $H =
H(F)$.  The latter relation contains the crucial information about the
underlying physical continuum (i.e., about spacetime and/or about the
material medium). Mathematically, this constitutive law arises either
from a suitable phenomenological theory of a medium or from the
electromagnetic field Lagrangian. It can be a nonlinear or even
nonlocal relation between the electromagnetic excitation and the field
strength.

Earlier, we have investigated the propagation of waves in the most
general local and linear theory. One can conveniently split the
constitutive relation into the three irreducible pieces which are
known as the {\it principal\/} part, the {\it skewon\/} part and the
{\it axion\/} part. It is important to stress that such a
decomposition is intrinsically metricfree, and thus it suits nicely
for the purpose of the study of the violations of the Lorentz symmetry
in the photon sector.  More specifically, we investigated the
influence of the axion and the skewon fields on the wave propagation
by applying a linear response formalism, see \cite{Obukhov:2000nw}.
Subsequently, Itin \cite{Yakov2004} studied birefringence caused by
the axion field in the Caroll-Field-Jackiw model of electrodynamics
\cite{Carroll}.  More recently, we investigated the birefringence of
electromagnetic waves and its consequences for the metric of spacetime
\cite{L-H}. The possible limits of the Lorentz symmetry can be seen in
a particularly transparent way, both qualitatively and quantitatively,
including also the case of a possible violation of the charge
conservation law \cite{chargenon}.

The axion field and its interference with electrodynamics is a
well-studied subject, and the same applies to the influence of the
general principal part of the constitutive tensor which was thoroughly
analyzed in \cite{KosteleckyMewes02}. However, this is not so in the
case of the skewon field, a new structure which allows us to exhaust
all 36 parameters of a linear spacetime relation (exact definitions
are given below). The only relevant earlier results are those of
Nieves and Pal \cite{NP89,NP94} which were confined to the case of the
spatially isotropic and constant scalar skewon field. In contrast, we
present here some results which hold true for all electrodynamical
models with an arbitrary local and linear {\it spacetime relation}
that contains a nontrivial skewon part. We use the expression
``spacetime relation'', since it applies to spacetime (``the vacuum'')
itself, and in order to distinguish it from the constitutive law of a
material medium. From a mathematical point of view, the general local
and linear spacetime relation is similar to the constitutive relation
for matter with sufficiently complicated electric and magnetic
properties. Quite naturally, the study and the physical interpretation
of the wave propagation in a manifold with a general spacetime
relation appears to be very close to the analysis of the corresponding
wave effects in crystal optics. The latter subject was extensively
studied in the literature, see \cite{Szivessy,Rama,Schaefer,LL84}, for
example. In particular, our analysis reveals the analogy of the
effects of the real skewon field to the propagation of waves in absorbing
media with circular dichroism, whereas the imaginary skewon field
appears to be analogous to the optical activity tensor.

If local coordinates $x^i$ are given, with $i,j,... =0,1,2,3$, we can
decompose the excitation and field strength 2-forms into their
components according to
\begin{equation}
H = {\frac 1 2}\,H_{ij}\,dx^i\wedge dx^j,\qquad
F = {\frac 1 2}\,F_{ij}\,dx^i\wedge dx^j.\label{geo1}
\end{equation}

%%%%%%%%%%%%%%%%%%%%%%%%%%%%%%%%%%%%%%%%%%%%%%%%%%%%%%%%%%%%
\section{General local and linear constitutive relation}
%%%%%%%%%%%%%%%%%%%%%%%%%%%%%%%%%%%%%%%%%%%%%%%%%%%%%%%%%%%%

We confine ourselves to the case in which the electromagnetic
excitation and the field strength are related by the local and linear
constitutive law,
\begin{equation}\label{HchiF}
H_{ij}={\frac 1 2}\,\kappa_{ij}{}^{kl}\,F_{kl}\,.
\end{equation}
The constitutive tensor $\kappa$ has 36 independent components.  One
can decompose this object into its irreducible pieces. Obviously,
contraction is the only tool for such a decomposition. Following Post
\cite{Postmap}, we can define the contracted tensor of type $[^1_1]$
\begin{equation}
\kappa_i{}^k := \kappa_{il}{}^{kl}\,,
\end{equation}
with 16 independent components. 
The second contraction yields the pseudo-scalar function
\begin{equation}
\kappa := \kappa_k{}^k = \kappa_{kl}{}^{kl}\,.
\end{equation}
The traceless piece
\begin{equation}
\not\!\kappa_i{}^k := \kappa_i{}^k - {\frac 1 4}\,\kappa\,\delta_i^k
\end{equation}
has 15 independent components. These pieces can now be subtracted out
from the original constitutive tensor. Then,
\begin{eqnarray}
  \kappa_{ij}{}^{kl} &=& {}^{(1)}\kappa_{ij}{}^{kl} +
  {}^{(2)}\kappa_{ij}{}^{kl} + {}^{(3)}\kappa_{ij}{}^{kl} \\ &=&
  {}^{(1)}\kappa_{ij}{}^{kl} +
  2\!\not\!\kappa_{[i}{}^{[k}\,\delta_{j]}^{l]} + {\frac 1
    6}\,\kappa\,\delta_{[i}^k\delta_{j]}^l.\label{kap-dec}
\end{eqnarray}
By construction, ${}^{(1)}\kappa_{ij}{}^{kl}$ is the totally traceless
part of the constitutive tensor:
\begin{equation}
{}^{(1)}\kappa_{il}{}^{kl} = 0.\label{notrace}
\end{equation}
Thus, we split $\kappa$ according to $36 = 20 + 15 + 1$, and the
$[^2_2]$ tensor ${}^{(1)}\kappa_{ij}{}^{kl}$ is subject to the 16
constraints (\ref{notrace}) and carries $20 = 36 -16$ components.

One may call ${}^{(1)}\kappa_{ij}{}^{kl}$ the principal, or the
metric-dilaton part of the constitutive law. Without such a term,
electromagnetic waves are ruled out, see
\cite{nonsym32,mexmeet,dublin}. We further identify the two other
irreducible parts with a {\it skewon} and an {\it axion} field,
respectively. Conventionally, the skewon and the axion fields are
introduced by
\begin{equation}
\!\not\!S_i{}^j = -\,{\frac 1 2}\!\not\!\kappa_i{}^j,\qquad
\alpha = {\frac 1 {12}}\,\kappa.\label{Salpha}
\end{equation}

The standard Maxwell-Lorentz electrodynamics arises when both skewon and
axion vanish, whereas 
\begin{equation}
{}^{(1)}\kappa_{ij}{}^{kl} = \lambda_0\,\eta_{ij}{}^{kl}.\label{ML}
\end{equation}
Here $\lambda_0 = \sqrt{\varepsilon_0/\mu_0}$ is the vacuum impedance.
A spacetime metric $g_{ij}$ is assumed on the manifold, and with $g :=
{\rm det}g_{ij}$ one defines $\eta_{ijkl} := \sqrt{-g}
\hat{\epsilon}_{ijkl}$ and $\eta_{ij}{}^{kl} =
\eta_{ijmn}\,g^{mk}g^{nl}$.  It has been shown recently \cite{L-H}
that taking the linear spacetime relation for granted, one ends up at
a Riemannian lightcone provided one forbids birefringence in vacuum,
see also \cite{Laem}.

Along with the original $\kappa$-tensor, it is convenient to introduce 
an alternative representation of the constitutive tensor:
\begin{equation}
\chi^{ijkl} := {\frac 1 2}\,\epsilon^{ijmn}\,\kappa_{mn}{}^{kl}.\label{chikap}
\end{equation}
Substituting (\ref{kap-dec}) into (\ref{chikap}), we find the corresponding
decomposition
\begin{equation}
\chi^{ijkl} = {}^{(1)}\chi^{ijkl} + {}^{(2)}\chi^{ijkl}
+ {}^{(3)}\chi^{ijkl}\label{chi-dec}
\end{equation}
with the principal, skewon, and axion pieces defined by
\begin{eqnarray}
{}^{(1)}\chi^{ijkl} &=& {\frac 1 2}\,\epsilon^{ijmn}\,\,{}^{(1)}
\kappa_{mn}{}^{kl},\label{chi1}\\ 
{}^{(2)}\chi^{ijkl} &=& {\frac 1 2}\,\epsilon^{ijmn}\,\,{}^{(2)}
\kappa_{mn}{}^{kl} = -\,\epsilon^{ijm[k}\!\not\!\kappa_m{}^{l]},\label{chi2}\\
{}^{(3)}\chi^{ijkl} &=& {\frac 1 2}\,\epsilon^{ijmn}\,\,{}^{(3)}
\kappa_{mn}{}^{kl} = {\frac 1 {12}}\,\epsilon^{ijkl}\,\kappa. \label{chi3}
\end{eqnarray}

Using the S-identity and the K-identity derived in \cite{mexmeet}, we
can verify that $^{(2)}\chi$ is {\it skew-symmetric} under the
exchange of the first and the second index pair, whereas $^{(1)}\chi$
is {\it symmetric}:
\begin{equation}
{}^{(2)}\chi^{ijkl} = -\,{}^{(2)}\chi^{klij},\qquad
{}^{(1)}\chi^{ijkl} = {}^{(1)}\chi^{klij}.
\end{equation}

\subsection{Space-time decomposed constitutive relation}

Making a $(1+3)$-decomposition \cite{HO02} of covariant
electrodynamics, we can write $H$ and $F$ as column 6-vectors with the
components built from the magnetic and electric excitation 3-vectors
${\cal H}_a, {\cal D}^a$ and electric and magnetic field strengths
$E_a, B^a$, respectively. Then the linear spacetime relation
(\ref{HchiF}) reads:
\begin{equation}
  \left(\begin{array}{c} {\cal H}_a \\ {\cal D}^a\end{array}\right) 
= \left(\begin{array}{cc} {{\cal C}}^{b}{}_a & {{\cal B}}_{ba} \\ 
{{\cal A}}^{ba}& {{\cal D}}_{b}{}^a \end{array}\right) \left(
\begin{array}{c} -E_b\\  {B}^b\end{array}\right)\,.\label{CR'}
\end{equation}
Here the constitutive tensor is conveniently represented by the 6-matrix
\begin{equation}\label{kappachi}
  \kappa_I{}^K=\left(\begin{array}{cc} {{\cal C}}^{b}{}_a & {{\cal
          B}}_{ba} \\ {{\cal A}}^{ba}& {{\cal D}}_{b}{}^a
    \end{array}\right)\,,\qquad \chi^{IK}= \left( \begin{array}{cc}
      {\cal B}_{ab}& {\cal D}_a{}^b \\ {\cal C}^a{}_b & {\cal A}^{ab}
    \end{array}\right)\,.
\end{equation}
The constitutive $3\times 3$ matrices ${\cal A,B,C,D}$ are constructed 
from the components of the original constitutive tensor as
\begin{eqnarray}\label{AB-matrix0}
{\cal A}^{ba}&:=& \chi^{0a0b}\,,\qquad
{\cal B}_{ba} := \frac{1}{4}\,\hat\epsilon_{acd}\,
\hat\epsilon_{bef} \,\chi^{cdef}\,,\\
\label{CD-matrix0}
{\cal C}^a{}_b& :=&\frac{1}{2}\,\hat\epsilon_{bcd}\,\chi^{cd0a}\,,\qquad
{\cal D}_a{}^b := \frac{1}{2}\,\hat\epsilon_{acd}
\,\chi^{0bcd}\,.
\end{eqnarray}
If we resolve with respect to $\chi$, we find the inverse formulas
\begin{eqnarray}\label{AB-matrix0'}
\chi^{0a0b} &=& {\cal A}^{ba}\,,\qquad 
\chi^{abcd} = \epsilon^{abe}\,\epsilon^{cdf}\,{\cal B}_{fe}\,,\\ 
\chi^{0abc} &=& \epsilon^{bcd}\,{\cal D}_d^{\ a}\,,\qquad
\chi^{ab0c} = \epsilon^{abd}\,{\cal C}^c_{\  d}\,.\label{CD-matrix0'}
\end{eqnarray}

The contributions of the principal, the skewon, and the axion parts to 
the above constitutive 3-matrices can be written explicitly as
\begin{eqnarray}
  {\cal A}^{ab} &=& -\varepsilon^{ab} -
  \epsilon^{abc}\!\not\!S_c{}^0,\label{A}\\ {\cal B}_{ab} &=&\>\;
  \mu_{ab}^{-1} + \hat{\epsilon}_{abc}\!\not\!S_0{}^c,\label{B}\\ 
  {\cal C}^a{}_b &=&\>\; \gamma^a{}_b\, - (\!\not\!S_b{}^a - \delta_b^a
  \!\not\!S_c{}^c) + \alpha\,\delta_b^a,\label{C}\\ {\cal D}_a{}^b &=&\>\;
  \gamma^b{}_a\, + (\!\not\!S_a{}^b - \delta_a^b \!\not\!S_c{}^c) +
  \alpha\,\delta_a^b. \label{D}
\end{eqnarray}
The set of the symmetric matrices $\varepsilon^{ab}=\varepsilon^{ba}$
and $\mu_{ab}^{-1} = \mu_{ba}^{-1}$ together with the traceless matrix
$\gamma^a{}_b$ (i.e., $\gamma^c{}_c =0$) comprise the principal part
${}^{(1)}\chi^{ijkl}$ of the constitutive tensor.  Usually,
$\varepsilon^{ab}$ is called {\it permittivity\/} tensor and
$\mu^{-1}_{ab}$ {\it reciprocal permeability\/} tensor
(``impermeability'' tensor), since they describe the polarization and
the magnetization of a medium, respectively. The magnetoelectric
cross-term $\gamma^a{}_b$ is related to the Fresnel-Fizeau effects.
The skewon contributions in (\ref{A}) and (\ref{B}) are responsible
for the electric and magnetic Faraday effects, respectively, whereas
skewon terms in (\ref{C}) and (\ref{D}) describe optical activity.

%%%%%%%%%%%%%%%%%%%%%%%%%%%%%%%%%%%%%%%%%%%%%%%%%%%%%%%%%%
\section{General Fresnel equations: wave and ray surfaces}
%%%%%%%%%%%%%%%%%%%%%%%%%%%%%%%%%%%%%%%%%%%%%%%%%%%%%%%%%%

Here we briefly summarize the results of previous work
\cite{Obukhov:2000nw,nonlwave,Dr.Guillermo,nonsym32}. In the Hadamard
approach, one studies the propagation of a discontinuity in the first
derivative of the electromagnetic field.  The basic notions are then
the fields of the {\it wave\/} covector and the {\it ray\/} vector
that encode the information about the propagation of a wave in a
spacetime with a general constitutive relation.

\subsection{Wave surface}

The crucial observation about the surface of discontinuity $S$
(defined locally by a function $\Phi$ such that $\Phi= const$ on $S$)
is that across $S$ the geometric Hadamard conditions are satisfied for
the components of the electromagnetic field and their derivatives:
$[F_{ij}] = 0, \,[\partial_i F_{jk}] = q_i\,f_{jk},\,[H_{ij}] =
0,\,[\partial_i H_{jk}] = q_i\, h_{jk}$. Here $q_i:=\partial_i\Phi$ is
the wave covector. Then using the Maxwell equations (\ref{me}) and the
constitutive law (\ref{HchiF}), we find a system of algebraic
equations for the jump functions:
\begin{equation}\label{4Dwave1}
  {\chi}^{\,ijkl}\, q_{j}\,f_{kl}=0 \,,\qquad {\epsilon}^{\,ijkl}\,
  q_{j}\,f_{kl}=0\,.
\end{equation}

Solving the last equation in (\ref{4Dwave1}) by means of $f_{ij} =
q_ia_j - q_ja_i$, we finally reduce (\ref{4Dwave1})$_1$ to
${\chi}{}^{\,ijkl}\,q_{j}q_ka_l=0$.  This algebraic system has a
nontrivial solution for $a_i$ only if the determinant of the matrix on
the left hand side vanishes. The latter gives rise to our {\it
  generalized covariant Fresnel equation}
\begin{equation} \label{Fresnel}  
{\cal G}^{ijkl}(\chi)\,q_i q_j q_k q_l = 0 \,,
\end{equation}
with the fourth order Tamm-Rubilar (TR) tensor density of weight $+1$
defined by
\begin{equation}\label{G4}  
  {\cal G}^{ijkl}(\chi):=\frac{1}{4!}\,\hat{\epsilon}_{mnpq}\,
  \hat{\epsilon}_{rstu}\, {\chi}^{mnr(i}\, {\chi}^{j|ps|k}\,
  {\chi}^{l)qtu }\,.
\end{equation}
It is totally symmetric, ${\cal G}^{ijkl}(\chi)= {\cal G}^{(ijkl)}(\chi)$,
and thus has  35 independent components.

Different irreducible parts of the constitutive tensor (\ref{kap-dec})
contribute differently to the general Fresnel equation. A
straightforward analysis \cite{mexmeet,dublin} shows that the axion
piece drops out completely from the TR-tensor, whereas the two
remaining irreducible parts of the constitutive tensor contributes to
(\ref{G4}) as follows:
\begin{equation} \label{propg8}
{\cal G}^{ijkl}(\chi) % = {\cal G}^{ijkl}({}^{(1)}\chi + {}^{(2)}\chi)
= {\cal G}^{ijkl}({}^{(1)}\chi) + {}^{(1)}
\chi^{\,m(i|n|j}\!\not\!S_m^{\ k} \!\not\!S_n^{\ l)}\,.
\end{equation}

\subsection{Ray surface}

A ray can be defined by a vector field $s= s^i\partial_i$ that is {\it
  dual} to the wave covector $q= q_i\,dx^i$ and to the wave front in the
following sense \cite{KiehnFresnel,Dr.Guillermo}:
\begin{equation}
s\rfloor h = 0,\qquad s\rfloor f = 0,\qquad s\rfloor q = 0.\label{svector}
\end{equation}
Here the 2-forms $h = {\frac 12}\,h_{ij}dx^i\wedge dx^j$ and $f =
{\frac 12} \,f_{ij}dx^i\wedge dx^j$ represent the jumps of the first
derivatives of the electromagnetic field.  Equations (\ref{svector})
are metric-free.  The system of the first two equations can be
analysed in complete analogy to the system (\ref{4Dwave1}). As a
result, one arrives at the ``dual" Fresnel equation imposed on the
components $s^i$ of the ray vector:
\begin{equation}
\widehat{\cal G}_{ijkl}\,s^is^js^ks^l = 0.\label{ray0} 
\end{equation}
Like the TR-tensor density (\ref{G4}), the totally symmetric tensor
density of weight $-1$ is constructed in terms of the components of the
constitutive tensor:
\begin{eqnarray}
\widehat{\cal G}_{ijkl}(\chi) &:=& {\frac 1{4!}}\,{\epsilon}^{mnpq}
\,{\epsilon}^{rstu}\,\widehat{\chi}_{mnr(i}\,\widehat{\chi}_{j|ps|k}
\,\widehat{\chi}_{l)qtu}\nonumber\\
&=& \frac{1}{4\cdot 4!}\,{\chi}^{pqmn}\,\hat{\epsilon}_{mnr(i}
\,\hat{\epsilon}_{j|pvw}\,{\chi}^{vwgh}\,\hat{\epsilon}_{ghs|k}
\,\hat{\epsilon}_{l)qtu}\,{\chi}^{turs}\,.\label{RG}
\end{eqnarray}
As before, a complete symmetrization has to be performed over the four indices 
$i,j,k,l$, with the vertical lines separating out those indices that are 
excluded from the symmetrization. % The extra factor $1/4$ is conventional. 
Here we introduced a ``double dual" of the constitutive tensor density
as
\begin{equation}
\label{lower}
  \widehat{\chi}_{ijkl}={\frac 14}\,\hat{\epsilon}_{ijmn}\,
  \hat{\epsilon}_{klpq}\,\chi^{mnpq}\,.
\end{equation}
The contraction of this new object with the original constitutive
tensor yields
\begin{equation}
{\frac 12}\,\widehat{\chi}_{klmn}\,\chi^{mnij} = {\frac 12}
\,\kappa_{kl}{}^{mn}\,\kappa_{mn}{}^{ij}.\label{chichi}
\end{equation}
Using (\ref{chi-dec}), we find the corresponding decomposition
\begin{equation}
  \widehat{\chi}_{ijkl}={}^{(1)}\widehat{\chi}_{ijkl} -
  2\hat{\epsilon}_{ijm[k} \,\!\not\!S_{l]}{}^m +
  \hat{\epsilon}_{ijkl}\,\alpha,\label{low-dec}
\end{equation}
Substituting this into (\ref{RG}), we arrive at the skewon
contribution to the ray surface:
\begin{equation}\label{rayskew}
  \widehat{\cal G}_{ijkl}(\chi) = \widehat{\cal G}_{ijkl}({}^{(1)}
  \chi) + {}^{(1)}\widehat\chi_{\,m(i|n|j}\!\not\!S_k{}^m
  \!\not\!S_{l)}{}^n\,.
\end{equation}

%%%%%%%%%%%%%%%%%%%%%%%%%%%%%%%%%%%%%%%%%%%%%%%%%%%%%%%%%%%%%%%%
\section{Skewon field: different parametrizations and wave propagation 
effects in vacuum}\label{effects}
%%%%%%%%%%%%%%%%%%%%%%%%%%%%%%%%%%%%%%%%%%%%%%%%%%%%%%%%%%%%%%%%

The main aim of this paper is to investigate the possible influence of the 
skewon field $\!\not\!S_i{}^j$ on the electromagnetic wave propagation.
For simplicity, we assume that the principal part of the constitutive
tensor is determined by the Maxwell-Lorentz law (\ref{ML}), that is,
\begin{equation}\label{chi0}
{}^{(1)}\chi^{minj} = \lambda_0\,\sqrt{-g}\left(g^{mn}g^{ji} 
- g^{mj}g^{ni}\right),\qquad {}^{(1)}\widehat{\chi}_{minj} = -\,{\frac 
{\lambda_0}{\sqrt{-g}}}\left(g_{mn}g_{ji} - g_{mj}g_{ni}\right).
\end{equation}
However, the skewon structure will be kept as general as possible. As
a preliminary step, we note that direct computations yield
\begin{equation}\label{TR1}
{\cal G}^{ijkl}({}^{(1)}\chi) = -\,\lambda_0^3\sqrt{-g}\,g^{(ij}g^{kl)},\qquad
\widehat{\cal G}_{ijkl}({}^{(1)}\chi) = {\frac {\lambda_0^3}{\sqrt{-g}}}
\,g_{(ij}g_{kl)}\,. 
\end{equation}

\subsection{`Double vector' parametrization}

The skewon, as a traceless tensor field of type $\left[^1_1 \right]$,
has 15 independent components. Given the {\it spacetimes metric}
$g_{ij}$, we can decompose the skewon field into an antisymmetric and
a symmetric part.  The latter can be conveniently constructed from a
pair of arbitrary (co)vector fields $v^i$ and $w_i\,$, so that we find
eventually
\begin{equation}
\!\not\!S_i{}^j = a_i{}^j + w_iv^j - {\frac 14}\,\delta_i^j\,(wv).\label{Svec}
\end{equation}
Here $a_{ij} := a_i{}^kg_{kj}= - a_{ji}$ is an arbitrary antisymmetric tensor 
and $(wv) = w_iv^i$. Such a parametrization produces almost a general skewon
field, since the number of independent components is here $14 = 6 (a_{ij}) 
+ 4 (v^i) + 4 (w_i)$.

Substituting (\ref{Svec}) into (\ref{propg8}) and (\ref{rayskew}), and 
taking into account (\ref{chi0}), we find 
\begin{eqnarray}
{}^{(1)}\chi^{\,m(i|n|j}\!\not\!S_m^{\ k} \!\not\!S_n^{\ l)} &=& \lambda_0
\,\sqrt{-g}\left[\left(a^{m(i}a_m{}^j + 2v^{(i}b^j + w^2\,v^{(i}v^j\right)
g^{kl)} - v^{(i}v^jw^kw^{l)}\right],\label{addTR1}\\ \label{addTR2}
{}^{(1)}\widehat\chi_{\,m(i|n|j}\!\not\!S_k{}^m \!\not\!S_{l)}{}^n &=& 
-\,{\frac {\lambda_0}{\sqrt{-g}}}\left[\left(a_{m(i}a^m{}_j + 2w_{(i}
\widehat{b}_j + v^2\,w_{(i}w_j\right)g_{kl)} - v_{(i}v_jw_kw_{l)}\right].
\end{eqnarray}
Here $b^i := w^j\,a_j{}^i$, $\,\widehat{b}_i := a_i{}^j\,v_j$ and 
$w^2 = w_iw^i,\; v^2 = v_iv^i$. 

Adding the principal terms (\ref{TR1}) to the contributions of the
skewon (\ref{addTR1}) and (\ref{addTR2}), we see that neither the wave
surface nor the ray surface decompose into the product of cones in
general.  In other words, birefringence is absent. The last terms in
(\ref{addTR1}) and in (\ref{addTR2}) are responsible for that. Let us
investigate those particular cases of the skewon field in the `double
vector' representation (\ref{Svec}) which do admit the birefringence
phenomenon.

\subsubsection{Purely antisymmetric skewon}\label{antiS}

When the skewon field is represented by its antisymmetric part only,
i.e., $\!\not\!\!S_i{}^j$ $= a_i{}^j$ (which is obtained by putting
both $v^i = w_i =0$), then the analysis of (\ref{propg8}) shows that the
light propagation is birefringent with one light cone defined by the
spacetime metric $g^{ij}$ and the second ``optical" metric
\begin{equation}
g_{(2)}^{ij} = g^{ij} - {\frac 1 {\lambda_0^2}}\,a^{mi}a_m{}^j.\label{opt1}
\end{equation}
This result actually holds true for any other parametrizations of the
skewon field which will be considered later. At the same time,
inspection of (\ref{rayskew}) yields the decomposition of the ray
surface into the light cone of the spacetime metric $g_{ij}$ and a
second cone defined by
\begin{equation}
g^{(2)}_{ij} = g_{ij} - {\frac 1 {\lambda_0^2}}\,a_{mi}a^m{}_j.\label{opt2}
\end{equation}
Although (\ref{opt1}) and (\ref{opt2}) look very similar, we notice
that they are not mutual inverse, and hence, they do not determine a
unique geometric structure on spacetime.

\subsubsection{Purely symmetric `vector' skewon}

Assume now that the skew-symmetric skewon part is absent, which is
evidently achieved if $a_{ij} = v_{[i}w_{j]}$. Then $b^i =
(1/2)[w^i(vw) - v^iw^2]$, $\,\widehat{b}_i = (1/2)[v_i(vw) - w_iv^2]$,
and we find the TR-tensor in the birefringent form
\begin{equation}
{\cal G}^{ijkl} = -\,\lambda_0^3\,\sqrt{-g}\,g_+^{(ij}g_-^{kl)}.\label{Gpm}
\end{equation}
Here the two light cones are defined by the two optical metrics
\begin{equation}
g_\pm^{ij} = g^{ij} - {\frac {v^{(i}w^{j)}}{2\lambda_0^2}}
\left[(vw) \pm \sqrt{(vw)^2 - 4\lambda_0^2}\right].\label{ometpm}
\end{equation}
Here $(vw) = v_iw^i$.  Obviously, in order to have the wave
propagation along separate light cones, the skewon field must satisfy
the condition
\begin{equation}
(vw)^2 \geq 4\lambda_0^2.\label{u2}
\end{equation}
When this condition is violated, the Fresnel wave surface is of 4th
order and birefringence is absent. For the equality sign in
(\ref{u2}), the birefringence disappears and the two light cones
coincide. However, one can straightforwardly prove that for $(vw)^2 =
4\lambda_0^2$ the optical metric has an Euclidean signature. Hence
there is no wave propagation in this case. The type of the signature
correlates with the sign of the determinant of the metric: the latter
is negative for the Lorentzian case and positive for the Euclidean
one.

A direct computation yields
\begin{equation}
\det\left(g^{ij} - kv^{(i}w^{j)}\right) = (\det g^{ij})\left\{1 -
k\,(vw) + {\frac {k^2}4}\left[(vw)^2 - v^2w^2\right]\right\}.\label{det}
\end{equation}
Comparing this with (\ref{ometpm}), we identify the coefficient as
$k_\pm = \left[(vw) \pm \sqrt{(vw)^2 -
    4\lambda_0^2}\right]/2\lambda_0^2$ and finally obtain
\begin{equation}
(\det g^{ij}_\pm) = (\det g^{ij})\,{\frac {k^2_\pm}4}\left[
(vw)^2 - v^2w^2 - 4\lambda_0^2\right].\label{detpm}
\end{equation}
Since the spacetime metric is Lorentzian, its determinant is negative,
$(\det g^{ij}) < 0$. As a result, we conclude that the optical metrics
will be also Lorentzian only if
\begin{equation}
(vw)^2 - v^2w^2 > 4\lambda_0^2.\label{u3}
\end{equation}
Note that the two conditions (\ref{u3}) and (\ref{u2}) look pretty
similar. In fact, we can verify that only one of them is an essential
inequality. Indeed, when the two vectors are either {\it both}
timelike ($v^2 > 0, w^2 > 0$) or {\it both\/} spacelike ($v^2 < 0, w^2
< 0$), the new condition (\ref{u3}) makes the constraint (\ref{u2})
redundant, whereas when at least one of the vectors is null,
(\ref{u3}) and (\ref{u2}) coincide.  Finally, when one of the vectors
is timelike and another spacelike, the only essential condition is
(\ref{u2}) and (\ref{u3}) is redundant.

With these conditions, we can classify the effects of the skewon field
on wave propagation. The spatially isotropic skewon field of Nieves \&
Pal \cite{NP89,NP94}, for example, belongs to the subcase when the two
vectors $w$ and $v$ are proportional to each other. Indeed, let us
consider the (flat) Minkowski spacetime with the Lorentzian metric
$g_{ij} = {\rm diag}(c^2,-1,-1,-1)$, and let us put $w^i = -\,v^i =
\delta^i_0 \,\sqrt{2S}/c$ with some scalar function $S$ of dimension
$[\lambda_0]$. Then, from (\ref{Svec}), we recover the isotropic
skewon field
\begin{equation}
\!\not\!S_i{}^j = \frac{S}{2}\left(\begin{array}{rccc} -3 & 0 & 0 & 0 \\ 0 &
1 & 0 & 0 \\ 0 & 0 & 1 & 0 \\ 0 & 0 & 0 & 1\end{array}\right).\label{PN}
\end{equation}
As a result, the spacetime relation becomes
\begin{eqnarray}\label{resultx}
{\cal D}^a &=&\;\, \varepsilon_0\,\delta^{ab} E_b\; +\left(- S 
+ \alpha\right)B^a\,, \\ {\cal H}_a &=& \left(- S - \alpha\right)E_a
+\mu_0^{-1}\,\delta_{ab}\, B^b\,.
\end{eqnarray}
Accordingly, in the special case when skewon and axion become {\it constant} 
fields, one can speak of the 4 electromagnetic constants for a vacuum 
spacetime with spatial isotropy: The electric constant $\varepsilon_0$, 
the magnetic constant $\mu_0$, the spatially isotropic part $S$ of the 
skewon $\!\not\!S_i{}^j$, and the axion $\alpha$. 

In this special case we evidently find $(vw)^2 - v^2w^2 = 0$. Hence
(\ref{detpm}) yields that both optical metrics have a wrong Euclidean
signature. This is in complete agreement with our earlier observations
\cite{mexmeet} in which the absence of wave propagation was reported
on the background of the isotropic skewon field of Nieves \& Pal.

However, now we obviously obtain a far more general result: whereas
Nieves \& Pal have considered only a very special configuration with
the spacelike vectors $v$ and $w$ being proportional to each other,
the wave propagation is actually also absent for any $w^i \sim v^i$
and any Riemannian spacetime metric $g_{ij}$. Then again $(vw)^2 -
v^2w^2 = 0$ and hence $(\det g^{ij}_\pm) = -\,(\det g^{ij})
\lambda_0^2k^2_\pm > 0$.

\subsection{General parametrization}

It is impossible to construct a general parametrization of the skewon
field by using covariant 4-dimensional objects. Instead, we have to
switch to the space and time decomposition techniques and to allow for
the use of 3-dimensional (3D) tensor objects. Then we can write the
required parametrization as follows:
\begin{equation}
  \!\not\!S_i{}^j = \left(\begin{array}{cc} -s_c{}^c & m^a \\ n_b &
      s_b{}^a\end{array}\right).\label{Sgen}
\end{equation}
Here $m^a$ and $n_b$ are 3D (co)vector fields, whereas $s_b{}^a$ is a
3D tensor of type $[_1^1]$. From now on, the indices from the
beginning of the Latin alphabet denote the spatial 3D components,
i.e., $a, b, c, \dots = 1,2,3$. The count of the independent
components, namely $3 + 3 + 9 = 15$, shows that we indeed have the
most general representation of the skewon field. It is straightforward
to establish the dimensions of the different pieces of skewon field:
We find that $[s_b{}^a] = [\lambda_0]$, $[n_b] = [\varepsilon_0]$ and
$[m^a] = 1/[\mu_0]$.

With this general parametrization, the linear spacetime relation (\ref{CR'})
can be written symbolically as follows:
\begin{equation}
\left(\begin{array}{c} {\cal H}_a \\ {\cal D}^a\end{array}\right) =
\left(\begin{array}{c}\hspace{-8pt}\null\\ \hspace{-8pt}\null
\end{array}\right)\hspace{-15pt}\kappa \;\,
 \left(\begin{array}{c} -E_b\\ {B}^b\end{array}\right)\,.
 \end{equation}
Here the linear operator on the right-hand side explicitly reads:
\begin{equation}
  \left(\begin{array}{c}\hspace{-8pt}\null\\ \hspace{-8pt}\null
\end{array}\right)\hspace{-15pt}\kappa \;\, :=
\left(\begin{array}{cc} \gamma^b{}_a & \mu_{ab}^{-1}  \\
       -\varepsilon^{ab}  & \gamma^a{}_b \end{array}\right)
 + \left(\begin{array}{cc}  - s_a{}^b +\delta_a^b
  s_c{}^c &\>\;
   - \hat{\epsilon}_{abc}m^c \\
  \epsilon^{abc}n_c &
  s_b{}^a - \delta_b^a s_c{}^c
  \end{array}\right)
   +\alpha\left(\begin{array}{cc}\delta_a^b&0\\0&\delta_b^a
\end{array}\right)
\,.\label{CR''''}
\end{equation}

Now we again specialize to the Maxwell-Lorentz principal part
(\ref{chi0}).  Moreover, since we are specifically interested in the
skewon effects, we will assume for concreteness that the spacetime
metric is the {\it flat Minkowskian} one with Lorentzian signature,
$g_{ij} = (c^2, -\, \delta_{ab}) = {\rm diag}(c^2,-1,-1,-1)$. The
Euclidean 3-metric tensor $\delta_{ab}$ should not be confused with
the Kronecker delta: The components of the latter are always either 0
or 1, whereas the former has these values in Cartesian coordinates
only. Then one can further decompose the 3D skewon tensor into its
symmetric and skewsymmetric parts:
\begin{equation}
s_b{}^a = u_b{}^a + \epsilon_b{}^{ac}z_c.\label{3tensor}
\end{equation}
Here $u_{ab} = u_{ba}$, and the 3D indices are raised and lowered by
means of the Euclidean 3D metric, i.e., $u_b{}^a =: u_{bc}\delta^{ac}$
and $\epsilon_b{}^{ac} := \epsilon_{bde}\delta^{ad}\delta^{ce}$. The
covector field $z_c$ describes the antisymmetric part of the 3D tensor
skewon.  Obviously, both irreducible parts of the tensor skewon have
the same dimension $[u_b{}^a] = [z_c] = [\lambda_0]$.

\subsubsection{Space-time decomposition and wave and ray surfaces}

The analysis of the wave propagation in the presence of the general
skewon field (\ref{Sgen}) is not possible in the above covariant
4-dimensional framework. Since the space and time components are
explicitly separated in the general parametrization (\ref{Sgen}), we
need a corresponding approach to the wave and ray surfaces in which a
space and time decomposition has been performed.

Namely, let us denote the independent components of the TR-tensor 
(\ref{G4}) as follows:
\begin{eqnarray}  
 \label{ma0} M &:=& {\cal G}^{0000} = \det{\cal A} \,,\\
M^a &:=& 4\,{\cal G}^{000a} = -\hat{\epsilon}_{bcd}\left( {\cal A}^{ba}
\,{\cal A}^{ce}\,{\cal C}^d_{\ e} + {\cal A}^{ab}\,{\cal A}^{ec}
\,{\cal D}_e^{\ d}\right)\,,\label{ma1}\\
 M^{ab} &:=& 6\,{\cal G}^{00ab} = \frac{1}{2}\,{\cal A}^{(ab)}\left[
({\cal C}^d{}_d)^2 + ({\cal D}_c{}^c)^2 - ({\cal C}^c{}_d + {\cal D}_d{}^c)
({\cal C}^d{}_c +  {\cal D}_c{}^d)\right]\nonumber\\ 
&&\qquad + ({\cal C}^d{}_c + {\cal D}_c{}^d)({\cal A}^{c(a}{\cal C}^{b)}{}_d 
+ {\cal D}_d{}^{(a}{\cal A}^{b)c}) - {\cal C}^d{}_d{\cal A}^{c(a}{\cal C}^{b)}
{}_c\nonumber\\ &&\qquad - {\cal D}_c{}^{(a}{\cal A}^{b)c}{\cal D}_d{}^d -
  {\cal A}^{dc}{\cal C}^{(a}{}_c {\cal D}_d{}^{b)} + 
  \left({\cal A}^{(ab)}{\cal A}^{dc}- {\cal A}^{d(a}{\cal A}^{b)c}\right)
   {\cal B}_{dc}\,,\label{ma2}\\
M^{abc} &:=& 4\,{\cal G}^{0abc} =
  \epsilon^{de(c|}\left[{\cal B}_{df}( {\cal A}^{ab)}\,{\cal D}_e^{\ f} 
 - {\cal D}_e^{\ a}{\cal A}^{b)f}\,) \right. \nonumber \\
&&\qquad\left.+ {\cal B}_{fd}({\cal A}^{ab)}\,{\cal C}_{\ e}^f - {\cal A}^{f|a}
{\cal C}_{\ e}^{b)}) +{\cal C}^{a}_{\ f}
\,{\cal D}_e^{\ b)}\,{\cal D}_d^{\ f} + {\cal D}_f^{\ a}
\,{\cal C}^{b)}_{\ e}\,{\cal C}^{f}_{\ d} \right] \,,\label{ma3}\\
M^{abcd} &:=& {\cal G}^{abcd} = \epsilon^{ef(c}\epsilon^{|gh|d}\,{\cal B}_{hf}
  \left[\frac{1}{2} \,{\cal A}^{ab)}\,{\cal B}_{ge} - {\cal C}^{a}_{\ 
      e}\,{\cal D}_g^{\ b)}\right] \,.\label{ma4}
\end{eqnarray}
Then, in decomposed form, the generalized Fresnel equation
(\ref{Fresnel}) reads
\begin{equation}
  q_0^4 M + q_0^3q_a\,M^a + q_0^2q_a q_b\,M^{ab} + q_0q_a q_b
  q_c\,M^{abc} + q_a q_b q_c q_d\,M^{abcd}=0\,.\label{decomp}
\end{equation} 
%%%%%%%%%%%%%%%%%%%%%%%%%%%%%% FIGURE %%%%%%%%%%%%%%%%%%%%%%%%%%%%%%%%%%
\begin{figure}
\begin{center}
\resizebox{8 cm} {!}
   {\includegraphics{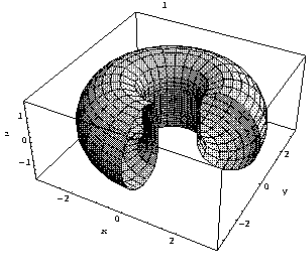}}
\caption{{\it Fresnel surface for a real skewon of the
electric Faraday type. It has the form of a toroid (depicted with a cut).
Here $\overline{n}_a = \overline{n}\,\delta_a^3$ with $\overline{n}^2 = 10$;
we use the dimensionless variables $x := cq_1/q_0,\, y := cq_2/q_0,\, z :=
cq_3/q_0$.}}\label{eFarfig1}
\end{center}
\end{figure}
%%%%%%%%%%%%%%%%%%%%%%%%%%% END FIGURE %%%%%%%%%%%%%%%%%%%%%%%%%%%%%%%%%%
Similarly to (\ref{ma0})-(\ref{ma4}), we can decompose the
4-dimensional tensor density (\ref{RG}) into a set of 3-dimensional
objects:
\begin{eqnarray}  
\label{mh0} \widehat{M} &:=& \widehat{\cal G}_{0000} = \det{\cal B} \,,\\
\widehat{M}_a &:=& 4\,\widehat{\cal G}_{000a} = -\epsilon^{bcd}\left( 
{\cal B}_{ba}\,{\cal B}_{ce}\,{\cal D}_d{}^e + {\cal B}_{ab}\,{\cal B}_{ec}
\,{\cal C}^e{}_d\right)\,,\label{mh1}\\
 \widehat{M}_{ab} &:=& 6\,\widehat{\cal G}_{00ab} = \frac{1}{2}
\,{\cal B}_{(ab)}\left[({\cal C}^d{}_d)^2 + ({\cal D}_c{}^c)^2 
- ({\cal C}^c{}_d + {\cal D}_d{}^c)({\cal C}^d{}_c +  {\cal D}_c{}^d)\right]
\nonumber\\ 
&&\qquad + ({\cal C}^c{}_d + {\cal D}_d{}^c)({\cal B}_{c(a}{\cal D}_{b)}{}^d 
+ {\cal C}^d{}_{(a}{\cal B}_{b)c}) - {\cal D}_d{}^d{\cal B}_{c(a}
{\cal D}_{b)}{}^c\nonumber\\ 
&&\qquad - {\cal C}^c{}_{(a}{\cal B}_{b)c}{\cal C}^d{}_d - {\cal B}_{dc}
{\cal D}_{(a}{}^c {\cal C}^d{}_{b)} + \left({\cal B}_{(ab)}{\cal B}_{dc}
- {\cal B}_{d(a}{\cal B}_{b)c}\right){\cal A}^{dc}\,,\label{mh2}\\
\widehat{M}_{abc} &:=& 4\,\widehat{\cal G}_{0abc} = \hat{\epsilon}_{de(c}
\left[{\cal A}^{df}( {\cal B}_{ab)}\,{\cal C}^e{}_f 
 - {\cal C}^e{}_a{\cal B}_{b)f}\,) \right. \nonumber \\
&&\qquad\left.+ {\cal A}^{fd}({\cal B}_{ab)}\,{\cal D}_f{}^e - {\cal B}_{|f|a}
{\cal D}_{b)}{}^e) + {\cal D}_{a}{}^f\,{\cal C}^e{}_{b)}\,{\cal C}^d{}_f 
+ {\cal C}^f{}_a\,{\cal D}_{b)}{}^e\,{\cal D}_f{}^d\right] \,,\label{mh3}\\
\widehat{M}_{abcd} &:=& \widehat{\cal G}_{abcd} = \hat{\epsilon}_{ef(c}
\hat{\epsilon}_{|gh|d}\,{\cal A}^{hf}\left[\frac{1}{2}\,{\cal B}_{ab)}
\,{\cal A}^{ge} - {\cal D}_a{}^e\,{\cal C}^g{}_{b)}\right]\,.\label{mh4}
\end{eqnarray}
Then the Fresnel type equation for the quartic ray surface (\ref{ray0}) 
in decomposed form reads
\begin{equation}
(s^0)^4\, \widehat{M} + (s^0)^3\,s^a\,\widehat{M}_a + (s^0)^2\,s^as^b
\,\widehat{M}_{ab} + s^0s^a s^b s^c\,\widehat{M}_{abc} 
+ s^a s^b s^c s^d\,\widehat{M}_{abcd}=0\,.\label{decompray}
\end{equation} 
Now we are in a position to study the effects induced by the different
irreducible parts of the skewon field.

%%%%%%%%%%%%%%%%%%%%%%%%%%%%%% FIGURE %%%%%%%%%%%%%%%%%%%%%%%%%%%%%%%%%%
\begin{figure}
\begin{center}
\resizebox{8 cm} {!}
   {\includegraphics{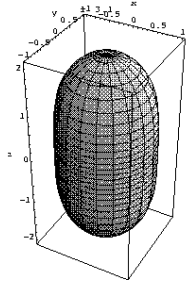}}
\caption{{\it Fresnel surface for a skewon of the
electric Faraday type. It has the form of a spheroid for the large purely
imaginary skewon $n$. Here $\overline{n}_a = \overline{n}\,\delta_a^3$ with
$\overline{n}^2 = -10$; we use the dimensionless variables $x := cq_1/q_0,
\, y := cq_2/q_0,\, z := cq_3/q_0$.}}\label{eFarfig2}
\end{center}
\end{figure}
%%%%%%%%%%%%%%%%%%%%%%%%%%% END FIGURE %%%%%%%%%%%%%%%%%%%%%%%%%%%%%%%%%%

\subsubsection{Skewonic electric Faraday effect}

When only the $n_a$ part of the skewon is nontrivial, whereas $m^a =
0$ and $s_b{}^a = 0$, the electric constitutive relation reads ${\cal
  D}^a = \varepsilon_0\,E^a + \epsilon^{abc}n_bE_c$. We will call this
the skewonic electric Faraday effect, since it describes the electric
Faraday effect in optically active media for purely imaginary $n_a$
\cite{Post,LL84,BornWolf}. Then, from (\ref{ma0})-(\ref{ma4}), we find
the coefficients of the Fresnel equation as $M^a = 0$, $M^{abc} = 0$,
whereas the nontrivial components read
\begin{eqnarray}
M &=& -\,\varepsilon_0\left(\varepsilon_0^2 + n^2\right),\qquad 
M^{abcd} = -\,{\frac 1 {\mu_0}}\,\lambda_0^2\,\delta^{(ab}\delta^{cd)},\\
M^{ab} &=& \varepsilon_0\left[2\lambda_0^2\,\delta^{ab} + c^2
\left(\delta^{ab}\,n^2 - n^a\,n^b\right)\right].
\end{eqnarray}
Here, as usual, $n^a = \delta^{ab}n_b$ and $n^2 = n^an_a$. 

On the other hand, according to (\ref{mh0})-(\ref{mh4}), the analysis
of the ray surface yields $\widehat{M}_a = 0$, $\widehat{M}_{abc} =
0$, and
\begin{equation}
\widehat{M} = {\frac 1 {\mu_0^3}},\qquad \widehat{M}_{ab} = -\,{\frac
{2\lambda_0^2} {\mu_0}}\,\delta_{ab},\qquad \widehat{M}_{abcd} =
\varepsilon_0\delta_{(ab}\left(\lambda_0^2\,\delta_{cd)} + c^2\,n_cn_{d)}
\right). 
\end{equation}
As a result, we obtain the following Fresnel surfaces for the wave covectors
and the ray vectors, respectively:
\begin{eqnarray}
(q_0/c)^4\,(1 + \overline{n}^2) - (q_0/c)^2\left[2q^2 + \overline{n}^2q^2
- (\overline{n}q)^2\right] + (q^2)^2 &=& 0,\label{Fqn}\\ \label{Fsn}
(s_0c)^4 - 2(s_0c)^2\,s^2 + s^2\left[s^2 + (\overline{n}s)^2\right] &=& 0.
\end{eqnarray}
Here we denote $q^2 = q_aq^a$, $s^2 = s_as^a$, $(\overline{n}q) = 
\overline{n}_aq^a$, $(\overline{n}s) = \overline{n}_as^a$. Moreover, we use 
the dimensionless skewon vector $\overline{n}_a := n_a/\varepsilon_0$. 

With the help of the nontrivial vector $\overline{n}$, we can split
the wave covector $q_a = q_a^\bot + q_a^{||}$ into its transversal and
longitudinal parts, with $q_a^\bot = q_a -
\overline{n}^{-2}\,\overline{n}_a \overline{n}^bq_b$ and
$q_a^{||}=\overline{n}^{-2}\,\overline{n}_a\overline{n}^b q_b$. Then the
Fresnel equation (\ref{Fqn}) is rewritten in the form
\begin{equation}
\left[(q_0/c)^2 - q_\bot^2\right]\left[(q_0/c)^2(1 + \overline{n}^2) - 
q_\bot^2\right] + q_{||}^2\left[-2(q_0/c)^2 + 2q_\bot^2 + q_{||}^2\right] =0. 
\end{equation}
Accordingly, in the plane transversal to the skewon vector $n_a$, we
find birefringence with the Minkowski metric and the second optical
metric $g^{ij} = {\rm diag}((1 + \overline{n}^2)/c^2, -1, -1, -1)$.
However, the Fresnel surface does not decompose into a product of the
two light cones. Although, for $n^2 = 0$, the wave covector surface is
a sphere, it becomes a toroid already for a small positive $n^2 > 0$.
The latter looks like a very thin and narrow belt for small values of
$n^2$. However, it becomes a thick toroid with the inner radius
decreasing to zero for large values of $n^2$. A typical Fresnel
surface for this case is depicted in Fig.~\ref{eFarfig1}. Physically,
the hole of this toroid corresponds to the directions in which waves
cannot propagate: In spherical coordinates, the wave propagation
occurs for $\theta_0 \leq \theta \leq \pi - \theta_0$, where the
spherical angle $\theta$ is counted from the direction of the skewon
vector $n_a$ and the limiting angle is determined from
$\sin^{2}\theta_0 = 2/(1 + \sqrt{1 + \overline{n}^2})$. As we see, the
hole disappears when $n^2\rightarrow\infty$.

This qualitative characterization of the wave propagation can be
completed by a numerical estimate for the velocity of light. The phase
velocity $v$ of the wave propagation is defined by the components of
the wave covector according to $q_a = q_0k_a/v$, where $k_a$ are the
components of the 3D unit covector. Usually, the mean value of the
velocity is of interest that is obtained after averaging over the
directions of propagation and the polarizations. For the case under
consideration, a direct calculation yields
\begin{equation}
<\!v^2\!>\, =\, c^2\left({\frac {1 + \overline{n}^2/3}{1 + \overline{n}^2}}\right).
\end{equation}
Thus, since $\overline{n}^2$ is positive, the velocity of the wave
propagation turns out to be less than $c$. Making use of the data on
the gamma-ray bursts and following \cite{KosteleckyMewes02}, we derive
as an upper limit for the skewonic electric Faraday effect $n^2 <
3\times 10^{-27}\,\varepsilon_0^2$.
%%%%%%%%%%%%%%%%%%%%%%%%%%%%%% FIGURE %%%%%%%%%%%%%%%%%%%%%%%%%%%%%%%%%%
\begin{figure}
\begin{center}
\resizebox{8 cm} {!}
   {\includegraphics{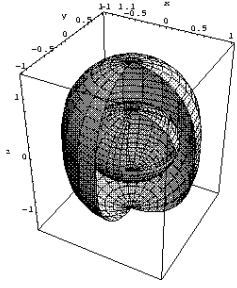}}
\caption{{\it
Fresnel surface for a skewon of the electric Faraday type. It has the form
of two concentric spheroids (depicted here with cuts) for a small purely
imaginary skewon $n$. Here $\overline{n}_a = \overline{n}\,\delta_a^3$ with
$\overline{n}^2 = -0.5$; we use the dimensionless variables $x := cq_1/q_0,
\,y := cq_2/q_0,\,z := cq_3/q_0$.}}\label{eFarfig3}
\end{center}
\end{figure}
%%%%%%%%%%%%%%%%%%%%%%%%%%% END FIGURE %%%%%%%%%%%%%%%%%%%%%%%%%%%%%%%%%%

The existence of a hole in the wave covector surface confirms our
earlier conclusion about the dissipative nature of the skewon. In
particular, when we continue to the purely imaginary values of $n$ so
that the square becomes negative, $n^2 < 0$, the topology of the
Fresnel surface changes drastically.  The hole disappears and the wave
covector surface assumes the form of a spheroid around the origin.
More exactly, there is a single spheroid elongated in the direction of
the skewon vector $n_a$ for $\overline{n}^2 \geq -1$, and two
(embedded one into another) spheroids when $-1 < \overline{n}^2 < 0$.
The average radius of the inner spheroid in the latter case is smaller
than 1 which, in physical terms, means that the velocity of wave
propagation is greater than $c$. Although the negative $n^2 < 0$
corresponds to the well known electric Faraday effect in crystals,
such a superluminal behavior of waves for $-1 < \overline{n}^2 < 0$
was never reported in the literature, to the best of our knowledge. In
Fig.~\ref{eFarfig2} and Fig.~\ref{eFarfig3}, we depict typical
Fresnel wave covector surfaces for $\overline{n}^2 = -10$ and
$\overline{n}^2 = -0.5$, respectively. The wave surface for
$\overline{n}^2=-1$ looks qualitatively like Fig.~\ref{eFarfig2}
with the top and the bottom of the spheroid at $z = \pm 1$.

\subsubsection{Skewonic magnetic Faraday effect}

In the opposite case, when the $m^a$ part of the skewon is nontrivial,
whereas $n_a = 0$ and $s_b{}^a = 0$, the magnetic constitutive
relation reads ${\cal H}_a = \mu_0^{-1}\,B_a + \epsilon_{abc}m^bB^c$.
We call this case the skewonic magnetic Faraday effect, since it
corresponds to the magnetic Faraday effect when $m^a$ is purely
imaginary \cite{Post,LL84,Polder}. From (\ref{ma0})-(\ref{ma4}), in
full analogy to the previous subsection,  we find that $M^a = 0$,
$M^{abc} = 0$, whereas the nontrivial components are
\begin{equation}
  M = -\,\varepsilon_0^3,\qquad M^{ab} = 2\varepsilon_0\lambda_0^2\,
  \delta^{ab}, \qquad M^{abcd} = -\,{\frac 1 {\mu_0}}\delta^{(ab}
  \left(\lambda_0^2 \,\delta^{cd)} + c^{-2}\,m^cm^{d)}\right).
\end{equation}
Analogously, from (\ref{mh0})-(\ref{mh4}), we find for the ray surface
that $\widehat{M}_a = 0$, $\widehat{M}_{abc} = 0$, and
 \begin{eqnarray}
\widehat{M} &=& {\frac 1 {\mu_0}}\left({\frac 1 {\mu_0^2}} + m^2\right),\qquad 
\widehat{M}_{abcd} = \varepsilon_0\,\lambda_0^2\,\delta_{(ab}\delta_{cd)},\\
\widehat{M}_{ab} &=& -\,{\frac 1 {\mu_0}}\left[2\lambda_0^2\,\delta_{ab} 
+ c^{-2}\left(\delta_{ab}\,m^2 - m_a\,m_b\right)\right].
\end{eqnarray}
Here again $m_a = \delta_{ab}m^b$ and $m^2 = m^am_a$. If, similarly to
the previous subsection, we introduce the dimensionless skewon vector
by $\overline{m}^a := m^a\mu_0$, we end up with the following Fresnel
wave and ray surfaces, respectively:
\begin{eqnarray}\label{Fqm}
(q_0/c)^4 - 2(q_0/c)^2\,q^2 + q^2\left[q^2 + (\overline{m}q)^2\right] &=& 0,\\
(s_0c)^4\,(1 + \overline{m}^2) - (s_0c)^2\left[2s^2 + \overline{m}^2s^2
- (\overline{m}s)^2\right] + (s^2)^2 &=& 0.\label{Fsm}
\end{eqnarray}
Similarly as before, $(\overline{m}q) = \overline{m}_aq^a$ and
$(\overline{m}s) = \overline{n}_as^a$.
%%%%%%%%%%%%%%%%%%%%%%%%%%%%%% FIGURE %%%%%%%%%%%%%%%%%%%%%%%%%%%%%%%%%%
\begin{figure}
\begin{center}
\resizebox{8 cm} {!}
   {\includegraphics{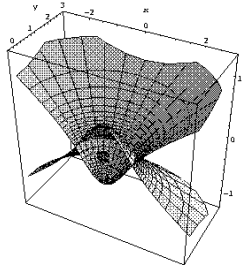}}
\caption{{\it
Fresnel surface for a skewon of the magnetic Faraday type. It has two branches
that are both hyperboloids for the large purely imaginary skewon $m$
(depicted with a cut into half). Here $\overline{m}^a = \overline{m}
\,\delta^a_3$ with $\overline{m}^2 = -10$; the dimensionless variables are 
$x := cq_1/q_0,\,y := cq_2/q_0,\,z := cq_3/q_0$.}}\label{mFarfig2}
\end{center}
\end{figure}
%%%%%%%%%%%%%%%%%%%%%%%%%%% END FIGURE %%%%%%%%%%%%%%%%%%%%%%%%%%%%%%%%%%
Repeating the calculation of the velocity of the wave propagation for
this case, we find the mean velocity $<\!v^2\!>\, =\, c^2$. As a
result, the analysis of the gamma-ray bursts data does not impose any
limit on the pure skewon field of the magnetic Faraday type.

\begin{figure}
\begin{center}
\resizebox{8 cm} {!}
   {\includegraphics{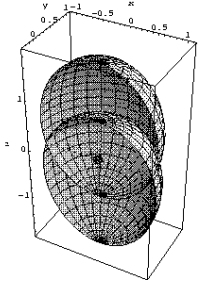}}
\caption{{\it
Fresnel surface for a skewon of the magnetic Faraday type. It has two branches
that are both spheroids for a small purely imaginary skewon $m$ (depicted with
a cut into half). Here $\overline{m}^a = \overline{m}\,\delta^a_3$ with
$\overline{m}^2 = -0.5$; the dimensionless variables are
$x := cq_1/q_0,\,y := cq_2/q_0,\,z := cq_3/q_0$.}}\label{mFarfig3}
\end{center}
\end{figure}

\begin{figure}
\begin{center}
\resizebox{8 cm} {!}
   {\includegraphics{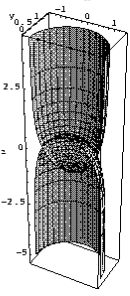}}
\caption{{\it Fresnel surface for a skewon of the magnetic Faraday type.
It has two branches that are both paraboloids for the purely imaginary
skewon with $\overline{m}^2 = -1$ (depicted with a cut into half). Here
$\overline{m}^a = \overline{m}\,\delta^a_3$; we use the dimensionless
variables $x := cq_1/q_0,\,y := cq_2/q_0,\,z := cq_3/q_0$.}}\label{mFarfig1}
\end{center}
\end{figure}

As one can see immediately, the wave covectors are only trivial in the
plane transverse to the direction of the skewon vector, i.e., when
$(\overline{m}q) = 0$. The waves propagate in that plane without
birefringence along the light cone defined by the standard Minkowski
metric, since then $[(q_0/c)^2 - q^2]^2 = 0$ is fulfilled. In
contrast, a purely imaginary skewon vector, with $m^2 < 0$, yields
more interesting results. In this case, the mentioned circle
represents a particular configuration that turns out to be the
intersection curve of the two branches of the Fresnel surface. The
typical Fresnel wave covector surfaces for the cases $\overline{m}^2 =
-10$ and $\overline{m}^2 = -0.5$ are depicted in Fig.~\ref{mFarfig2}
and Fig.~\ref{mFarfig3}, respectively: Both branches are
hyperboloids or spheroids, intersecting in the $z=0$ plane. The wave
surface for $\overline{m}^2 = -1$ is qualitatively different: It is
represented by paraboloids as shown in Fig.~\ref{mFarfig1}.
%%%%%%%%%%%%%%%%%%%%%%%%%%%%%% FIGURE %%%%%%%%%%%%%%%%%%%%%%%%%%%%%%%%%%
%%%%%%%%%%%%%%%%%%%%%%%%%%% END FIGURE %%%%%%%%%%%%%%%%%%%%%%%%%%%%%%%%%%

Comparison of the results of the two last subsections demonstrates
that the electric and magnetic Faraday skewon effects are dual to each
other in the following sense: When we interchange the wave and the ray
vectors $q \leftrightarrow s$ and simultaneously the skewon vectors
$\overline{n} \leftrightarrow\overline{m}$, then the wave surface
(\ref{Fqn}) has the same form as the ray surface (\ref{Fsm}) and vice
versa: The ray surface (\ref{Fsn}) has the same form as the wave
surface (\ref{Fqm}). In this sense,
Figs.~\ref{mFarfig2}-\ref{mFarfig1} show the ray (wave) surfaces
dual to the respective wave (ray) surfaces depicted in
Figs.~\ref{eFarfig1} and \ref{eFarfig2}, with the mentioned
replacement of $\overline{n} \leftrightarrow\overline{m}$ and with the
interchange between the dimensionless wave covector variables $x =
cq_1/q_0,\,y = cq_2/q_0,\,z = cq_3/q_0$ and the dimensionless ray
vector variables $x = s^1/s^0c,\,y = s^2/s^0c, \,z = s^3/s^0c$.

\subsubsection{Skewonic magneto-electric optical activity}

Let us now assume that only the tensor part of the skewon is present
in (\ref{Sgen}), whereas $m^a = 0$ and $n_a = 0$. We call this a
skewon of the magneto-electric type, since it corresponds to the case
of the natural optical activity in matter for the purely imaginary
$s_b{}^a$, see \cite{Post,LL84,Eyring,Feigel}.  Furthermore, we
consider the effects of the symmetric and skew-symmetric parts
(\ref{3tensor}). In the absence of the skew-symmetric tensor part ($z_c =
0$), we find $M^a = 0$, $M^{abc} = 0$ and
\begin{eqnarray}
M &=& -\,\varepsilon_0^3,\qquad M^{abcd} = {\frac 1 {\mu_0}}\left[
-\,\lambda_0^2\,\delta^{(ab}\delta^{cd)} + \delta^{(ab}u_e{}^cu^{|e|d)}
- u^{(ab}u^{cd)}\right],\\
M^{ab} &=& \varepsilon_0\left[2\lambda_0^2\,\delta^{ab} + 
\left(2u^{ab} - \delta^{ab}u_d{}^d\right)u_c{}^c - u_e{}^au^{|e|b)}\right].
\end{eqnarray}
As a check of the consistency of our formalism, we can verify that the
isotropic case $u_a{}^b = -\,S\delta_a{}^b$ reduces to the
birefringent case (\ref{Gpm}) and (\ref{ometpm}) with $w^i = -\,v^i =
\delta^i_0 \,\sqrt{2S}/c$.

As to the ray surface, a direct computation yields $\widehat{M}_a 
= 0$, $\widehat{M}_{abc} = 0$ and
\begin{equation}
\widehat{M} =\mu_0^{-3},\qquad \widehat{M}_{ab} = -\,c^2\,\delta_{aa'}
\delta_{bb'}\,M^{a'b'},\qquad \widehat{M}_{abcd} = -\,c^{-2}\,\delta_{aa'}
\delta_{bb'}\delta_{cc'}\delta_{dd'}\,M^{a'b'c'd'}.
\end{equation}
Accordingly, the form of the Fresnel ray and wave surfaces turns out
to be exactly the same.

A direct calculation of the mean velocity of the light propagation now
yields
\begin{equation}
<\!v^2\!>\, = c^2\left(1 - {\frac {\overline{u}_a{}^b\overline{u}_b{}^a 
+ \overline{u}_a{}^a\overline{u}_b{}^b} 6}\right).
\end{equation}
As usual, we again introduce the dimensionless skewon variable
$\overline{u}_a{}^b = u_a{}^b/\lambda_0$. Then, similarly to the above
cases, we find as upper value for the symmetric tensor skewon field
from gamma-ray burst data \cite{KosteleckyMewes02} $u_a{}^bu_b{}^a <
7\times 10^{-27}\,\lambda_0^2$.

\begin{figure}
\begin{center}
\resizebox{8 cm} {!}
   {\includegraphics{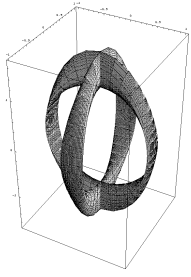}}
\caption{{\it Fresnel surface for a skewon of the magneto-electric optical 
activity type. It has two intersecting toroidal branches for the real skewon
$\overline{u} = 0.8$. We use the dimensionless
variables $x := cq_1/q_0,\,y := cq_2/q_0,\,z := cq_3/q_0$.}}\label{u08}
\end{center}
\end{figure}

The Fresnel wave covector surface has now quite a different form as
compared to the two cases above. For concreteness, let us take a
tensor skewon with only one nontrivial component, $u_a{}^b =
u\left(\delta_a^1\,\delta_2^b + \delta_a^2\,\delta_1^b\right)$. Then,
for small values of the skewon $\overline{u} < 1$, the Fresnel wave
surface is depicted in Fig.~\ref{u08}. As a comment to this figure, let
us recall that for a vanishing skewon $\overline{u} = 0$ we have a
pure vacuum spacetime relation, and the Fresnel wave surface is then a
{\it sphere}. With $\overline{u} \neq 0$, this sphere degenerates to a
pair of highly deformed intersecting toroids that, for extremely
small values of $\overline{u}$, are nearly covering the sphere. But
with growing $\overline{u}$, the toroids becomes thin and thinner. The
intermediate situation is actually depicted in Fig.~\ref{u08}.  Since
we do not expect large skewon fields in general, we here limit
ourselves to the case of a small $\overline{u}$.
%%%%%%%%%%%%%%%%%%%%%%%%%%%%%% FIGURE %%%%%%%%%%%%%%%%%%%%%%%%%%%%%%%%%%
%%%%%%%%%%%%%%%%%%%%%%%%%%% END FIGURE %%%%%%%%%%%%%%%%%%%%%%%%%%%%%%%%%%

In the complementary situation, when $m^a = 0$, $n_a = 0$ and the symmetric 
part is absent in (\ref{3tensor}), $u_a{}^b = 0$, we obtain a particular case 
described in Sec.~\ref{antiS}: Birefringence with the Minkowski light cone 
and the second optical metric
\begin{equation}\label{opt3}
g_{ij}^{(2)} = \left(\begin{array}{c|c} c^2 & 0 \\ \hline 0 & \left(
1 - z^2/\lambda_0^2\right)^{-1}\left( -\,\delta_{ab} + z_az_b/\lambda_0^2
\right)\end{array}\right).
\end{equation}
Unlike as with the symmetric tensor skewon, here the optical metric is
always Lorentzian with the determinant $(\det g_{ij}^{(2)}) =
-\,c^2/\left( 1 - z^2/\lambda_0^2\right)^2$. This means that the waves
propagates along both light cones (except for a skewon satisfying $z^2
= \delta^{ab} z_az_b = \lambda_0^2$ when the optical metric becomes
degenerate). A straightforward computation of the mean speed of the
wave propagation now yields
\begin{equation}
<\!v^2\!>\, = c^2\left(1 - {\frac {z^2} {3\lambda_0^2}}\right),
\end{equation}
and, accordingly, the upper limit for the skewon field from the gamma-ray
data is again $z^2 < 7\times 10^{-27}\lambda_0^2$. 

%%%%%%%%%%%%%%%%%%%%%%%%%%%%%%%%%%%%%%%%%%%%%%%%%%%%%%%%%%%%%%%%
\section{Skewon effects in matter}
%%%%%%%%%%%%%%%%%%%%%%%%%%%%%%%%%%%%%%%%%%%%%%%%%%%%%%%%%%%%%%%%

%%%%%%%%%%%%%%%%%%%%%%%%%%%%%% FIGURE %%%%%%%%%%%%%%%%%%%%%%%%%%%%%%%%%%
%%%%%%%%%%%%%%%%%%%%%%%%%%% END FIGURE %%%%%%%%%%%%%%%%%%%%%%%%%%%%%%%%%%
In the previous section we studied the wave propagation on a vacuum
spacetime described by the principal part ${}^{(1)}\chi^{ijkl}$ of the
constitutive tensor (\ref{chi0}). We found that the characteristic
effect of the skewon field in that case was the emergence of holes in
the wave covector surface.  In physical terms this means the complete
damping of the wave propagation in certain directions. Here we will
briefly study the wave propagation in an anisotropic dielectric medium
and demonstrate that a similar effect occurs in the presence of a
skewon field.

The constitutive relation for anisotropic matter is most easily
formulated in terms of the $3\times 3$-matrix parametrization
(\ref{kappachi})-(\ref{CD-matrix0}). In the {\it absence\/} of a
skewon field, an arbitrary dielectric medium is described by the
spacetime relation
\begin{equation}
{\cal A}^{ab} = -\,\varepsilon_0\,\varepsilon^{ab},\quad {\rm with}\quad
\varepsilon^{ab} = \left(\begin{array}{ccc}\varepsilon_1 & 0 & 0\\ 0& 
\varepsilon_2 & 0\\ 0& 0& \varepsilon_3\end{array}\right),\qquad 
{\cal  B}_{ab} = {\frac 1 {\mu_0}}\,\delta_{ab}.\label{ABani}
\end{equation}
The matrices ${\cal C}^a{}_b = 0$ and ${\cal D}_a{}^b = 0$. Then,
{}from (\ref{ma0})-(\ref{ma4}), we find $M^a =0$, $M^{abc} =0$, $M =
-\,\varepsilon_0^3\varepsilon_1\varepsilon_2\varepsilon_3$, $M^{abcd}
= -\,\delta^{(ab}\varepsilon^{cd)}\lambda_0^2/\mu_0$, and
\begin{equation}
M^{ab} = \varepsilon_0\lambda_0^2\left(\begin{array}{ccc} 
\varepsilon_1(\varepsilon_2 + \varepsilon_3)& 0 & 0\\
0 &\varepsilon_2(\varepsilon_1 + \varepsilon_3) & 0 \\ 0 & 0 &
\varepsilon_3(\varepsilon_1 + \varepsilon_2)\end{array}\right).\label{aniMab}
\end{equation}
As a result, the Fresnel equation for the wave surface (\ref{decomp})
can be recast into the simple form
\begin{equation}\label{wavesurf}
{\frac {\varepsilon_1\,q_1^2}{c^2 q^2 - \varepsilon_1\,q_0^2}} +
{\frac {\varepsilon_2\,q_2^2}{c^2 q^2 - \varepsilon_2\,q_0^2}} +
{\frac {\varepsilon_3\,q_3^2}{c^2 q^2 - \varepsilon_3\,q_0^2}} = 0;
\end{equation}
as in the previous section, $q^2 = q_aq^a$. 

\begin{figure}
\begin{center}
\resizebox{8 cm} {!}
   {\includegraphics{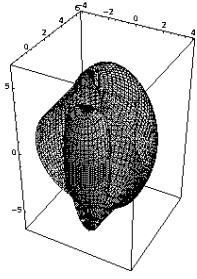}}
\caption{{\it Fresnel wave covector surface for an anisotropic dielectric
medium with $\varepsilon_1 = 39.7$, $\varepsilon_2 = 15.4$, $\varepsilon_3
= 2.3$. There are two branches, the outer part of the surface is cut into
half in order to show the inner branch; we use the dimensionless
variables $x := cq_1/q_0,\,y := cq_2/q_0,\,z := cq_3/q_0$.}}\label{aniW}
\end{center}
\end{figure}

\begin{figure}
\begin{center}
\resizebox{8 cm} {!}
   {\includegraphics{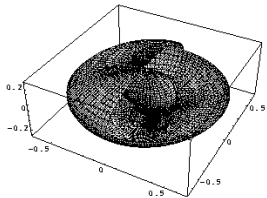}}
\caption{{\it Fresnel
ray vector surface for an anisotropic dielectric medium with $\varepsilon_1
= 39.7$, $\varepsilon_2 = 15.4$, $\varepsilon_3 = 2.3$. There are two
branches, 1/4 part of the outer surface is cut off in order to show the
second inner branch; we use the dimensionless variables
$x := s^1/s^0c,\,y := s^2/s^0c,\,z := s^3/s^0c$.}}\label{aniR}
\end{center}
\end{figure}

As to the ray surface, we can immediately verify {}from 
(\ref{mh0})-(\ref{mh4}) that $\widehat{M}_a =0$ and $\widehat{M}_{abc} =0$, 
whereas $\widehat{M} = \mu_0^{-3}$, $\widehat{M}_{abcd} = \varepsilon_0
\lambda_0^2\delta_{(ab}\varepsilon^{-1}{}_{cd)}\,\varepsilon_1\varepsilon_2
\varepsilon_3$, and
\begin{equation}
\widehat{M}_{ab} = -\,{\frac {\lambda_0^2}{\mu_0}}\left(\begin{array}{ccc}
\varepsilon_2 + \varepsilon_3 & 0 & 0\\ 0 & \varepsilon_1 + \varepsilon_3 
& 0 \\ 0 & 0 & \varepsilon_1 + \varepsilon_2\end{array}\right).\label{anihMab}
\end{equation}
Then, from (\ref{decompray}), one can straightforwardly derive
the dual equation of the Fresnel ray surface for the components of the
ray vector $s = \left(s^0, s^1, s^2, s^3\right)$:
\begin{equation}\label{raysurf}
{\frac {(s^1)^2}{\varepsilon_1\,s^2 - c^2(s^0)^2}} +
{\frac {(s^2)^2}{\varepsilon_2\,s^2 - c^2(s^0)^2}} +
{\frac {(s^3)^2}{\varepsilon_3\,s^2 - c^2(s^0)^2}} = 0.
\end{equation}
Typical wave and ray surfaces are depicted in Figs.~\ref{aniW} and
\ref{aniR}, respectively. Both surfaces consist of two
non-intersecting branches touching each other exactly in 4 points.
%%%%%%%%%%%%%%%%%%%%%%%%%%%%%% FIGURE %%%%%%%%%%%%%%%%%%%%%%%%%%%%%%%%%%
%%%%%%%%%%%%%%%%%%%%%%%%%%% END FIGURE %%%%%%%%%%%%%%%%%%%%%%%%%%%%%%%%%%

When the skewon field is present, the mentioned picture becomes more
nontrivial.  In particular, for the case of the isotropic skewon
(\ref{PN}), the constitutive relation is modified by adding to
(\ref{ABani}) the nonvanishing $3\times 3$ matrices
\begin{equation}\label{CDani}
{\cal C}^a{}_b = -\,S\delta^a_b,\qquad {\cal D}_a{}^b = S\delta_a^b.
\end{equation}
Then, a quick computation shows that all the $M$-coefficients remain the
same except that (\ref{aniMab}) is replaced with
\begin{equation}
M^{ab} = \varepsilon_0\lambda_0^2\left(\begin{array}{ccc} 
\varepsilon_1(\varepsilon_2 + \varepsilon_3 - 4\overline{S}^2)& 0 & 0\\
0 &\varepsilon_2(\varepsilon_1 + \varepsilon_3 - 4\overline{S}^2) & 0 \\ 
0 & 0 &\varepsilon_3(\varepsilon_1 + \varepsilon_2 - 4\overline{S}^2)
\end{array}\right).\label{aniMabS}
\end{equation}
Here we introduced the dimensionless skewon variable $\overline{S} 
:= S/\lambda_0$. 

Analogously we can verify that the dual $\widehat{M}$-coefficients are the
same as in the skewonless case except that (\ref{anihMab}) is changed to
\begin{equation}
\widehat{M}_{ab} = -\,{\frac {\lambda_0^2}{\mu_0}}\left(\begin{array}{ccc}
\varepsilon_2 + \varepsilon_3 - 4\overline{S}^2 & 0 & 0\\ 0 & \varepsilon_1 
+ \varepsilon_3  - 4\overline{S}^2 & 0 \\ 0 & 0 & \varepsilon_1 + 
\varepsilon_2 - 4\overline{S}^2\end{array}\right).\label{anihMabS}
\end{equation}
%%%%%%%%%%%%%%%%%%%%%%%%%%%%%% FIGURE %%%%%%%%%%%%%%%%%%%%%%%%%%%%%%%%%%
%%%%%%%%%%%%%%%%%%%%%%%%%%% END FIGURE %%%%%%%%%%%%%%%%%%%%%%%%%%%%%%%%%%

\begin{figure}
\begin{center}
\resizebox{8 cm} {!}
   {\includegraphics{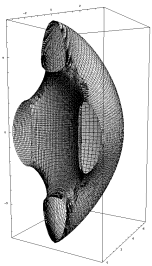}}
\caption{{\it Fresnel wave covector surface for an anisotropic dielectric
medium with $\varepsilon_1 = 39.7$, $\varepsilon_2 = 15.4$, $\varepsilon_3
= 2.3$ in presence of a skewon. The two original branches (cf. with
Fig.~\ref{aniW}) are now merged into a single surface with the wormholes
replacing the original four points where the branches touched. The surface
is cut into half; we use the dimensionless variables $x := cq_1/q_0,
\,y := cq_2/q_0,\,z := cq_3/q_0$.}}\label{holes}
\end{center}
\end{figure}

Then the equations of wave and ray surfaces cannot be represented any
longer in the compact form (\ref{wavesurf}) and (\ref{raysurf}). The
skewon affects the form of the surfaces by creating typical holes that
correspond to the directions of wave covectors along which no wave
propagation occurs.  More specifically, the original two branches of
the wave and ray surfaces, depicted in Figs.~\ref{aniW},\,\ref{aniR},
merge and form a single non-simply-connected surface. In the process
of such a merging, the four points where the original two branches
touched each other become the wormholes through which one can move
from the outer surface into the inner one.  Such a typical skewon
effect is shown in Fig.~\ref{holes}.
%%%%%%%%%%%%%%%%%%%%%%%%%%%%%% FIGURE %%%%%%%%%%%%%%%%%%%%%%%%%%%%%%%%%%
%%%%%%%%%%%%%%%%%%%%%%%%%%% END FIGURE %%%%%%%%%%%%%%%%%%%%%%%%%%%%%%%%%%

%%%%%%%%%%%%%%%%%%%%%%%%%%%%%%%%%%%%%%%%%%%%%%%%%%%%%%%%%%%%%%%%
\section{Discussion and conclusion}
%%%%%%%%%%%%%%%%%%%%%%%%%%%%%%%%%%%%%%%%%%%%%%%%%%%%%%%%%%%%%%%%

The violation of the Lorentz symmetry of spacetime results in
``spoiling" the ordinary light cone and, in particular, in the
emergence of birefringence in light propagation. Thus, the
investigation of electromagnetic wave phenomena provides an
understanding of the Lorentz violation in the photon sector of SME.
The birefringence type effects represent the observable consequences
of the model and the corresponding measurements impose limits on the
Lorentz-violating parameters in the general linear spacetime relation.
It was shown previously \cite{KosteleckyMewes02} by using cosmological
and laboratory observations that the birefringence-related parameters
of the principal part of the constitutive tensor must be smaller than
$10^{-32}\,\lambda_0$.

In this paper we investigated the general covariant Fresnel equation
for the wave covectors as well as for the ray vectors in the case of a
linear spacetime relation (\ref{HchiF}) that connects the
electromagnetic excitation with the field strength. Strictly speaking,
the physical information contained in the ray vector surface is the
same as that in the wave covector surface.  The reason why we still
analyzed the ray along with the wave surfaces is twofold. Firstly, we
developed the four-dimensional covariant approach to the ray surfaces
in complete analogy with the earlier results obtained for the Fresnel
wave surfaces. Such a formalism might be helpful in the study of the
various aspects of the wave propagation in media, since the
construction of the ray surfaces is quite a common tool in crystal
optics, see \cite{Szivessy,Rama,Schaefer,LL84,KiehnFresnel,BornWolf}.
Secondly, we demonstrated explicitly that although the ray surface is
dual to the wave surface (which is also clearly seen from the figures
above), the optical metric, which is naturally derived in the case of
birefringence, is different for the wave and for the ray surfaces and
the two metrics derived are {\it not} dual to each other.

The skewon part (\ref{Salpha})$_1$, (\ref{chi2}) of the constitutive
tensor was in the center of our study. We introduced its general
parametrization (\ref{Sgen}). This enabled us to distiguish between
three different types of effects: the electric and magnetic Faraday
effects and the (magneto-electric) optical activity. The influence of
a skewon field on the Fresnel wave and ray surfaces is qualitatively
different in these three case. This is illustrated in the figures in
the corresponding subsections of Sec.~\ref{effects}. However, the
characteristic sign of the skewon is the emergence of the specific
{\it holes in the Fresnel surfaces\/} that correspond to the
directions in space along which the wave propagation is damped out
completely. This effect is in complete agreement with our earlier
conclusion on the dissipative nature of the skewon field
\cite{HO02,nonsym32,mexmeet}.

Besides the qualitative results, one may derive numerical limits on
the magnitude of the skewon components from the experimental search
for the anisotropy of the velocity of light. Using the data for the
gamma-ray bursts and following \cite{KosteleckyMewes02}, we then
obtain the estimates $s_a{}^bs_b{}^a\sim a_a{}^ba_b{}^a < 7\times
10^{-27}\lambda_0^2\,$ and $n_an^a < 3\times 10^{-27}\varepsilon_0^2$.
This shows that if the skewon indeed spoils the light cone structure
(yielding a violation of the Lorentz symmetry and an anisotropy of the
propagation of light), its influence should be extremely small. As a
final remark we note that the experimental laboratory data
\cite{cavity} imposes much stronger constraints on the principal part
of the constitutive tensor (see the corresponding analysis in
\cite{KosteleckyMewes02}), and we can expect that the same will be
true for the skewon too since the theoretical formalisms are pretty
much the same in both cases.

\bigskip
{\bf Acknowledgment}. This work was supported by the Deutsche 
Forschungsgemeinschaft (Bonn) with the grant HE~528/20-1. The figures in
this paper were constructed with the help of a Mathematica program written
by Sergey Tertychniy (Moscow). We thank Sergey also for many discussions.

\end{document}